\documentclass[
  aps,
  prl,
  onecolumn,
  amsfonts,
  amssymb,
  amsmath,
  superscriptaddress,
  floatfix
]{revtex4-2}

\usepackage[utf8]{inputenc}
\usepackage{bm}  
\usepackage[normalem]{ulem}  
\usepackage{siunitx}
\usepackage{booktabs}
\usepackage{graphicx}
\usepackage[svgnames]{xcolor}
\usepackage{adjustbox}
\usepackage{tabularx}
\usepackage{xspace}  
\usepackage{comment}
\usepackage{tikz} 
\usepackage{import} 

\usepackage{hyperref}

\newcolumntype{Y}{>{\centering\arraybackslash}X}

\newcommand*{\vb}[1]{\boldsymbol{#1}}  
\newcommand*{\dd}{\mathrm{d}}  

\newcommand*{\laplacian}{\nabla^2}
\newcommand*{\gradient}{\vb{\nabla}}

\newcommand*{\uvec}{\vb{u}}
\newcommand*{\kvec}{\vb{k}}
\newcommand*{\xvec}{\vb{x}}

\newcommand*{\lvec}{\vb{l}}
\newcommand*{\rvec}{\vb{r}}
\newcommand*{\svec}{\vb{s}}

\newcommand*{\nablavec}{\boldsymbol{\nabla}}

\newcommand*{\Cloop}{\mathcal{C}}

\newcommand*{\CircR}{\Gamma_{\!r}}

\newcommand*{\lf}{\ell_{f}}
\newcommand*{\lalpha}{\ell_{\alpha}}

\newcommand*{\Lint}{\ell_\text{I}}  

\begin{document}

\title{Lack of self-similarity in transverse velocity increments and circulation statistics in two-dimensional turbulence}

\author{Nicol\'as P. M\"uller}
\affiliation{%
Laboratoire de Physique de l'École normale supérieure, ENS, Université PSL, CNRS, Sorbonne Université, Université Paris Cité, F-75005 Paris, France
}
\affiliation{%
Université Côte d'Azur, Observatoire de la Côte d'Azur, CNRS, Laboratoire Lagrange, Boulevard de l'Observatoire CS 34229 - F 06304 NICE Cedex 4, France
}
\author{Giorgio Krstulovic}
\affiliation{%
Université Côte d'Azur, Observatoire de la Côte d'Azur, CNRS, Laboratoire Lagrange, Boulevard de l'Observatoire CS 34229 - F 06304 NICE Cedex 4, France
}
\date{\today}

\begin{abstract}
  We study the statistics of longitudinal and transverse structure functions, as well as velocity circulation, in the inverse energy cascade of two-dimensional turbulence. Using direct numerical simulations of the incompressible Navier--Stokes equations, we show that transverse structure functions exhibit an anomalous scaling, in contrast to the self-similar behavior of longitudinal ones. 
   We derive an analytical relation that shows that the scaling exponents of transverse structure functions and velocity circulation are related in two-dimensional turbulence. 
\end{abstract}

\maketitle

Two-dimensional (2D) turbulence is a paradigmatic system in fluid dynamics as it strongly differs from three-dimensional (3D) turbulence. One of the most remarkable phenomenon which takes place in 2D turbulent flows is the inverse energy cascade, a process in which energy is transferred towards large scales leading to the formation of large-scale coherent structures \cite{Kraichnan1967,Leith1968,Batchelor1969}. This mechanism is relevant in geophysical flows such as atmospheres and oceans, which exhibit quasi-2D properties due to the suppression of motion in one direction induced by confinement, stratification, rotation, or other effects \cite{Davidson2013,Alexakis2018,Marino2013}. A prime example of this phenomenon is Jupiter's giant red spot \cite{Young2017}.

Unlike the direct energy cascade in three dimensions \cite{Frisch1995}, it has been widely accepted that the 2D inverse energy cascade is not intermittent. 
More precisely, most of the attention has been devoted to the longitudinal velocity increments  $\delta u^\parallel_r  = (\uvec(\xvec + \rvec) - \uvec(\xvec)) \cdot \hat{r}$, with $\uvec$ the velocity field, $\xvec$ the space coordinate and $\rvec$ the increment. This quantity has been shown to be self-similar for a given length scale $r$ within the inertial range \cite{Paret1998,Boffetta2000}.
It follows that the longitudinal structure functions (LSF) of order $p$ 
\begin{equation}
  S_p^\parallel(r) = \langle | \delta u^\parallel_r |^p \rangle\sim r^{\zeta_p^\parallel}
  \label{eq:Sp_long}
\end{equation}
exhibit scaling exponents $\zeta_p^\parallel$ that follow a self-similar dependence $p/3$, consistent with Kraichnan-Leith-Batchelor (KLB) theory \cite{Kraichnan1967,Leith1968,Batchelor1969}. The angular brackets $\langle . \rangle$ indicate averaging over space and time. Note that this is in stark contrast to 3D homogeneous and isotropic turbulence, where strong fluctuations in the velocity field lead to deviations from the Kolmogorov self-similar theory (K41) predicting $\zeta_p^\parallel=p/3$. Deviations from this scaling are typically described using multifractal theories \cite{Frisch1995}.  

In recent years, the study of the intermittent behavior in turbulent flows has taken a new direction through the analysis of velocity circulation, a conserved quantity in ideal fluids defined as the line integral of the velocity field around a closed loop $\Cloop_r$ of linear length scale $r$
\begin{equation}
  \CircR = \oint_{\Cloop_r} \uvec \cdot \dd\lvec.
  \label{eq:circulation_intro}
\end{equation}
High-resolution direct numerical simulations (DNS) of 3D turbulence showed an intermittent behavior in the statistics of velocity circulation, with scaling exponents $\lambda_p$ of a different nature than velocity increments \cite{Iyer2019}. That is, the naive relation between the scaling exponents of structure functions and circulation moments, expressed as $\lambda_p = \zeta_p + p$, is not satisfied. This behavior was also observed in quantum turbulence \cite{Muller2021,Polanco2021,Muller2022a}, where the nature of vortices drastically differs from that of classical fluids, motivating the development of new intermittency theories based on circulation \cite{Migdal2020, Apolinario2020, Moriconi2024}. A recent experimental study in quasi-2D turbulence showed that the statistics of velocity circulation is intermittent \cite{Zhu2023}, and was later confirmed in numerical simulations of 2D classical and quantum turbulence \cite{Muller2024}. This striking observation is in strong contrast with the established self-similar picture of 2D turbulent flows. An explanation for the surprising difference between fluctuations of circulation and velocity increments remains elusive.  

A priori, the relation between increments and circulation can be understood when writing explicitly the velocity circulation around a square loop of opposite vertices $(x_0,y_0)$ and $(x_1,y_1) = (x_0+r,y_0+r)$
\begin{eqnarray}
   \Gamma_r &=& \int_{x_0}^{x_0+r} \left[ u_x(x, y_0+r) - u_x(x, y_0) \right] \dd x +  \int_{y_0}^{y_0+r} \left[ u_y(x_0+r, y) - u_y(x_0,y) \right] \dd y.
  \label{eq:circulation_increments}
\end{eqnarray}
We note that the integrands correspond to the transverse velocity increments in different directions \cite{Supplemental}. 
The transverse scaling exponents $\zeta_p^\perp$ are defined from the transverse structure functions (TSFs) as
\begin{equation}
  S_p^\perp(r) = \langle | \delta \uvec^\perp_r |^p \rangle\sim r^{\zeta_p^\perp}
  \label{eq:Sp_trans},
\end{equation}
with $\delta \uvec^\perp_r = \delta \uvec_r - \delta u^\parallel_r \hat{r}$ the transverse velocity increments. It follows that a refined dimensional guess for the circulation scaling exponent would be $\lambda_p=\zeta_p^\perp + p$, but this relation does not hold in 3D turbulence.

Scaling exponents of LSFs and TSFs in 3D homogeneous and isotropic turbulence are known to differ, the latter being more intermittent \cite{Chen1997}. Furthermore, high-resolution DNS showed that the transverse scaling exponents $\zeta_p^\perp$ saturate for high-order moments \cite{Iyer2020}. In 2D, the TSFs have received much less attention.

In this Letter, we study the statistics of velocity increments and circulation in the inverse energy cascade of 2D turbulence by means of DNS. We look independently at the longitudinal, transverse and circulation structure functions and do a general comparison of their scaling properties. 
We also derive an analytical relation between the circulation and the transverse scaling exponents.

\begin{table}[b]
  \begin{tabularx}{.6\textwidth}{ X Y Y Y Y Y Y Y}
    RUNS & N & $\alpha$  & $\nu$ & $L / L_f $ & $\epsilon$ & $\Lint / L$ & $R_{\mathrm{I}}$ \\
    \hline
    RUN-A & 2048 & $0.02$   &  $10^{-5}$         & 256 & 0.001 & 0.135 & 10.61 \\
    RUN-B & 6144 & $0.0328$ &  $8\times 10^{-6}$ & 768 & 0.033 & 0.272 & 35.2\\
    \hline \hline
  \end{tabularx}
\caption{ Parameters of the 2D numerical simulations. $N$ is the number of linear collocation points, $\alpha$ the linear friction coefficient, $\nu$ the viscosity, $L/L_f$ the ratio between the system size and the forcing scale, $\epsilon$ the mean energy flux of the inverse cascade, $\Lint = \int k^{-1}E(k) \dd k / \int E(k) \dd k$ the integral scale, and $R_{\mathrm{I}} = (\Lint / L_f)^{2/3}$ a Reynolds number based on the integral and forcing scales. }
\label{tab:runs}
\end{table}

We describe the dynamics of a 2D flow using the incompressible Navier-Stokes equations for the scalar vorticity field $\omega = (\gradient \times \uvec) \cdot \hat{z}$ 
\begin{equation}
  \partial_t \omega + \left\{\omega, \psi\right\} = \nu \nabla^2 \omega - \alpha \omega + f,
  \label{eq:ns}
\end{equation}
where the vorticity and the stream function satisfy the relation $\omega = -\laplacian \psi$, with $\nu$ the kinematic viscosity and $f$ an external forcing. The Poisson brackets are defined as $\left\{\omega, \psi\right\} = \partial_x \omega \partial_y \psi - \partial_y \omega \partial_x \psi$, and the term $-\alpha \omega$ is a linear frictional damping that dissipates energy at large scales. 
We analyze data from two different DNS, one kindly provided by Guido Boffetta (RUN-A) with $N_x = N_y = N=2048$ linear collocation points, and a second one at a larger resolution performed by our group (RUN-B) with $N=6144$ \cite{Muller2024}. 
In both cases, Eq.~\eqref{eq:ns} is solved in a 2D periodic square of side $2\pi$ using a pseudo-spectral method and evolved using a second-order Adams-Bashforth (RUN-A) or Runge-Kutta (RUN-B) scheme. 
For RUN-A, the forcing is a random white noise in time with a Gaussian correlation in space, which satisfies $\langle f(\xvec,t) f(0, t') \rangle = F_0 l_f^2 \delta(t-t') \exp[-x^2/(2\ell_f^2)]$ with $\ell_f$ the forcing scale \cite{Boffetta2000}. For RUN-B, we use a constant in time Gaussian forcing in Fourier space $\hat{f}(\kvec) = \exp[-(|\kvec|-k_f)^2/ (2 \Delta k^2)]$, with $k_f = \ell_f^{-1}$ the forcing wave number and $\Delta k$ its width. We choose the parameters $\alpha$ and $k_f$ in order to maximize the width of the inertial range.
Due to the implementation of different forcings in each simulation, the statistical properties of the 2D turbulent flow slightly differ at the forcing length scale, but become negligible in the inertial range. 
The turbulent properties of the flow are averaged in time once the system reaches a statistically steady state. We average 28 fields for RUN-A and 453 fields for RUN-B. The parameters of each DNS are shown in Table~\ref{tab:runs}.

\begin{figure*}[h!]
  \centering
  \includegraphics[width=.99\textwidth]{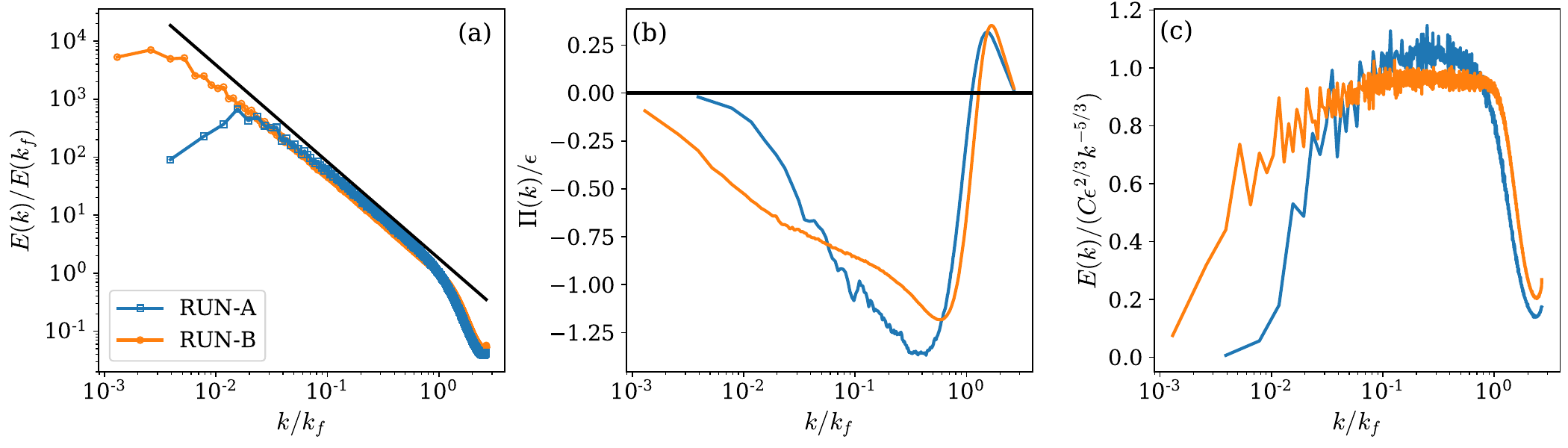}
  \caption[]{%
    (a) Energy spectra, (b) energy flux, and (c) compensated energy spectra for both DNS. The wave number is normalized by the forcing wave number $k_f$, and the flux is normalized by its mean value $\epsilon$. For the compensated spectra we used the constant $C=6$ \cite{Boffetta2012}.
    }
    \label{fig:spectra_and_flux}
\end{figure*}

The energy spectra for both runs develop a clear $k^{-5/3}$ scaling law between the large-scale dissipation wave number $k_\alpha \sim \ell_\alpha^{-1}$ with $\ell_\alpha \simeq \epsilon^{1/2} \alpha^{-3/2}$, and the forcing wave number $k_f$ (see Fig.~\ref{fig:spectra_and_flux}). 
In this same range of scales, the energy flux 
\begin{equation}
  \Pi(k) = \langle \uvec^{<k} \cdot \left[ \uvec \cdot \nablavec \uvec \right] \rangle
\end{equation}
takes negative values, signature of the energy cascading towards large scales. The superscript $^{<k}$ indicates that the velocity field considers only wave numbers such that $k < |\kvec|$ and $\epsilon = k_f^{-1} \sum_{k=1}^{k_f} \Pi(k)$ is the mean energy flux. 
The formation of a condensate at small wave numbers is avoided by the damping term.
Note that the energy flux is not exactly flat in the inertial range due to the use of a linear damping. It has been observed that employing an hypoviscosity term flattens the energy flux in the inertial range \cite{Boffetta2007}.
It is expected that in the limit of an infinite scale separation $\lalpha / \lf \rightarrow \infty$, the flux becomes constant in a larger range of scales. Figure~\ref{fig:spectra_and_flux} (c) shows the energy spectra compensated by the KLB prediction $C \epsilon^{2/3} k^{-5/3}$ with $C = 6$ \cite{Boffetta2012}.

\begin{figure*}[t]
  \centering
  \includegraphics[width=.99\textwidth]{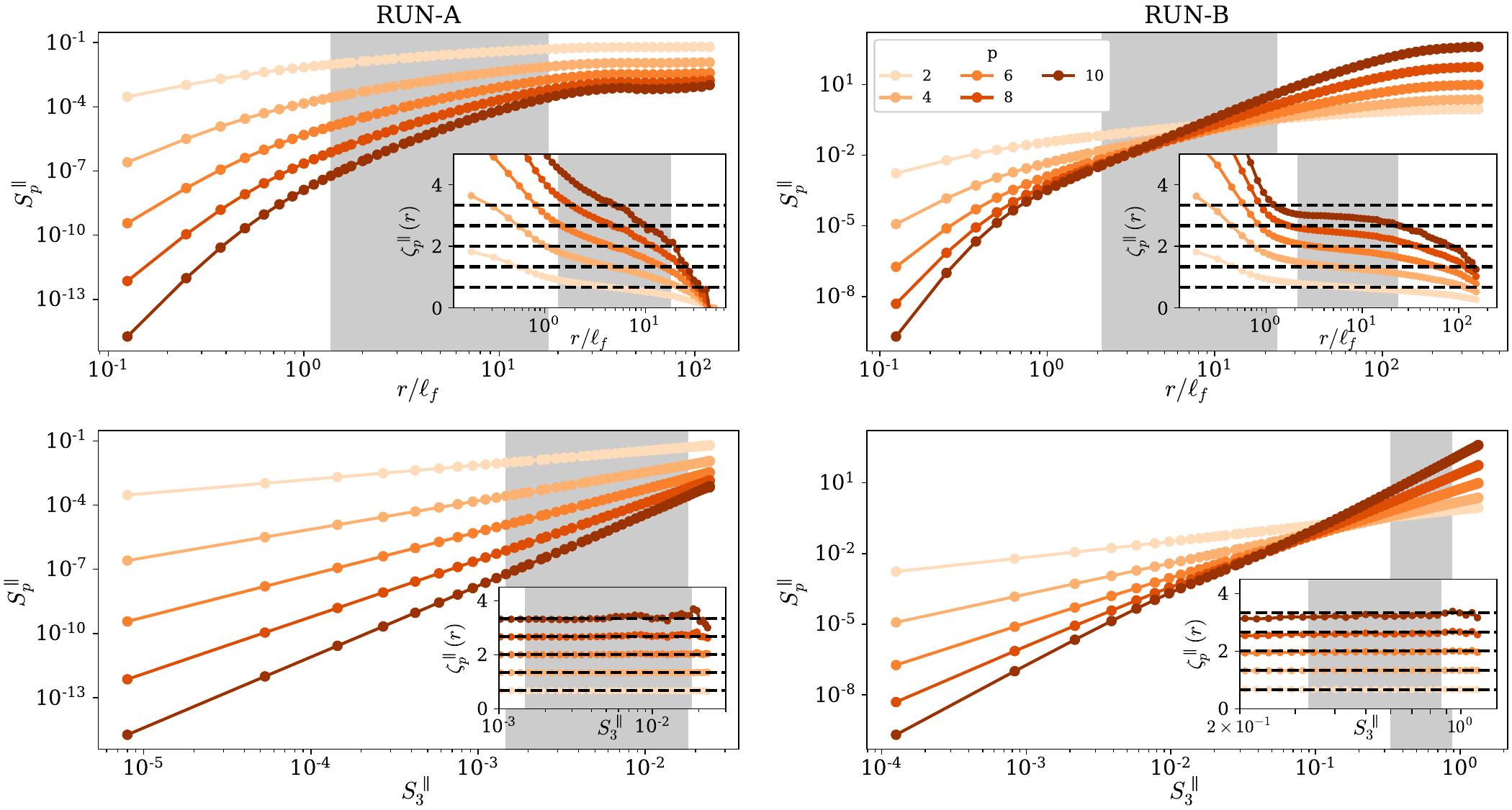}
  \caption[]{%
    Longitudinal structure functions in numerical simulations of 2D turbulence in the inverse energy cascade for moments up to $p=10$ for (left column) RUN-A and (right column) RUN-B. Top panels show the structure functions as a function of $r$, and the bottom panels show the ESS method. The shaded area indicates the inertial range. The insets show the local slopes $\zeta_p^{\parallel} = \dd \log S_p^{\parallel} / \dd \log r$. Dashed horizontal lines in the insets show the self-similar scaling $\zeta_p^{\mathrm{ss}} = p/3$.  
    }
    \label{fig:Sp_long}
\end{figure*}

For the analysis of structure functions, we compute the velocity increments along lines in both the $x$ and $y$ directions.  
The top row of Fig.~\ref{fig:Sp_long} shows the LSF defined as in Eq.~\eqref{eq:Sp_long} for even-order moments up to $p=10$ for both DNS. 
The insets display the local slopes, defined as the logarithmic derivative $\zeta_p^\parallel(r) = \dd \log S_p^\parallel / \dd \log r$. Note that for large scales $r > \lalpha$, the flow is decorrelated and the structure functions become flat, while for small scales $r < \lf$ they follow the viscous scaling $S_p \sim r^{2p}$. In the inertial range, the LSFs present logarithmic corrections which affect the scaling properties of the flow \cite{Paret1998}. The deviations from a pure power-law scaling are stronger for RUN-A than for RUN-B, probably due to the different resolutions used in each dataset, as well as the peculiarities of the forcing. 
To reduce these systematic deviations, we employ the extended self-similarity (ESS) method with respect to the third-order moment $S_3^\parallel(r) = \langle |\delta u^\parallel_r|^3 \rangle$ \cite{Benzi1993}. The shaded area indicates the inertial range $\ell_f < r < \ell_\alpha$ defined from the ESS method. 
Note that, due to the implementation of different forcings, the moments at the injection scale clearly differ from each other. Nevertheless, this effect in the inertial range is suppressed employing the ESS method, as shown in the bottom row of Fig.~\ref{fig:Sp_long}. The new local slopes defined as $\dd \log S_p / \dd \log S_3$ display clear scaling properties that follow the well-known self-similar behavior $\zeta_p^\parallel = p/3$ \cite{Boffetta2000}. 

\begin{figure*}[h!]
  \centering
  \includegraphics[width=.99\textwidth]{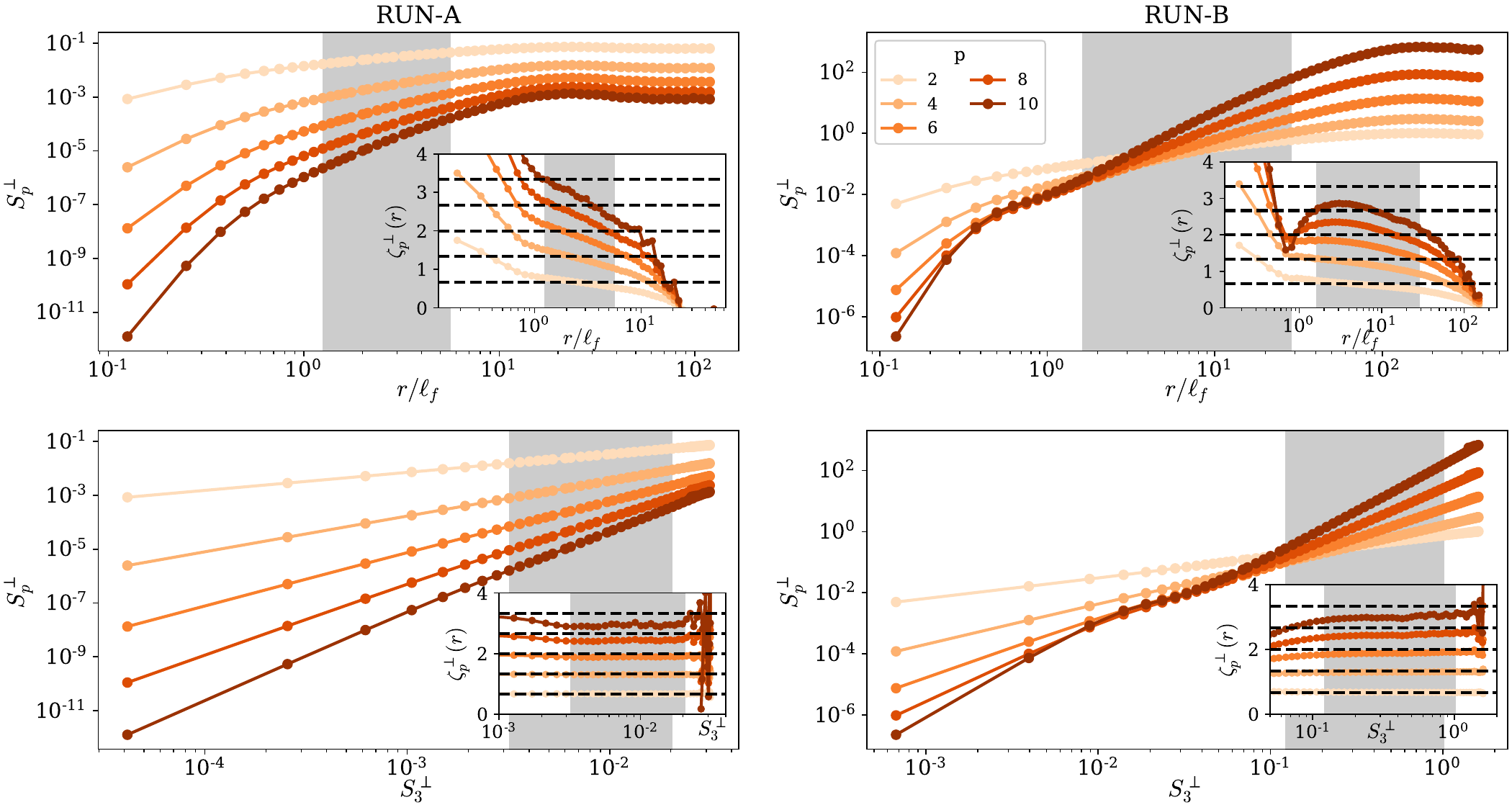}
  \caption[]{%
    Transverse structure functions up to $p=10$ in numerical simulations of 2D turbulence in the inverse energy cascade for (left column) RUN-A and (right column) RUN-B. Top panels show the structure functions as a function of $r$, and the bottom panels show the ESS method. The shaded area indicates the inertial range. The insets show the local slopes $\zeta_p^{\perp}(r) = \dd \log S_p^{\perp} / \dd \log r$. Dashed horizontal lines in the insets show the self-similar scaling $\zeta_p = p/3$.  
    }
    \label{fig:Sp_tran}
\end{figure*}
Following the same procedure, we compute the TSFs. Figure \ref{fig:Sp_tran} shows the TSFs for both datasets.
Again, the structure functions display logarithmic corrections in the inertial range that we reduce by employing the ESS method with respect to $S_3^\perp$. 
Strikingly, the local slopes deviate from the self-similar prediction, suggesting that fluctuations of transverse increments are stronger than longitudinal ones. This behavior is somehow similar to the case of 3D turbulence, where the $\zeta_p^\perp$ saturate for $p \geq 10$ \cite{Iyer2020}. We do not observe this saturation in 2D turbulence. 

\begin{figure*}[h!]
  \centering
  \includegraphics[width=.99\textwidth]{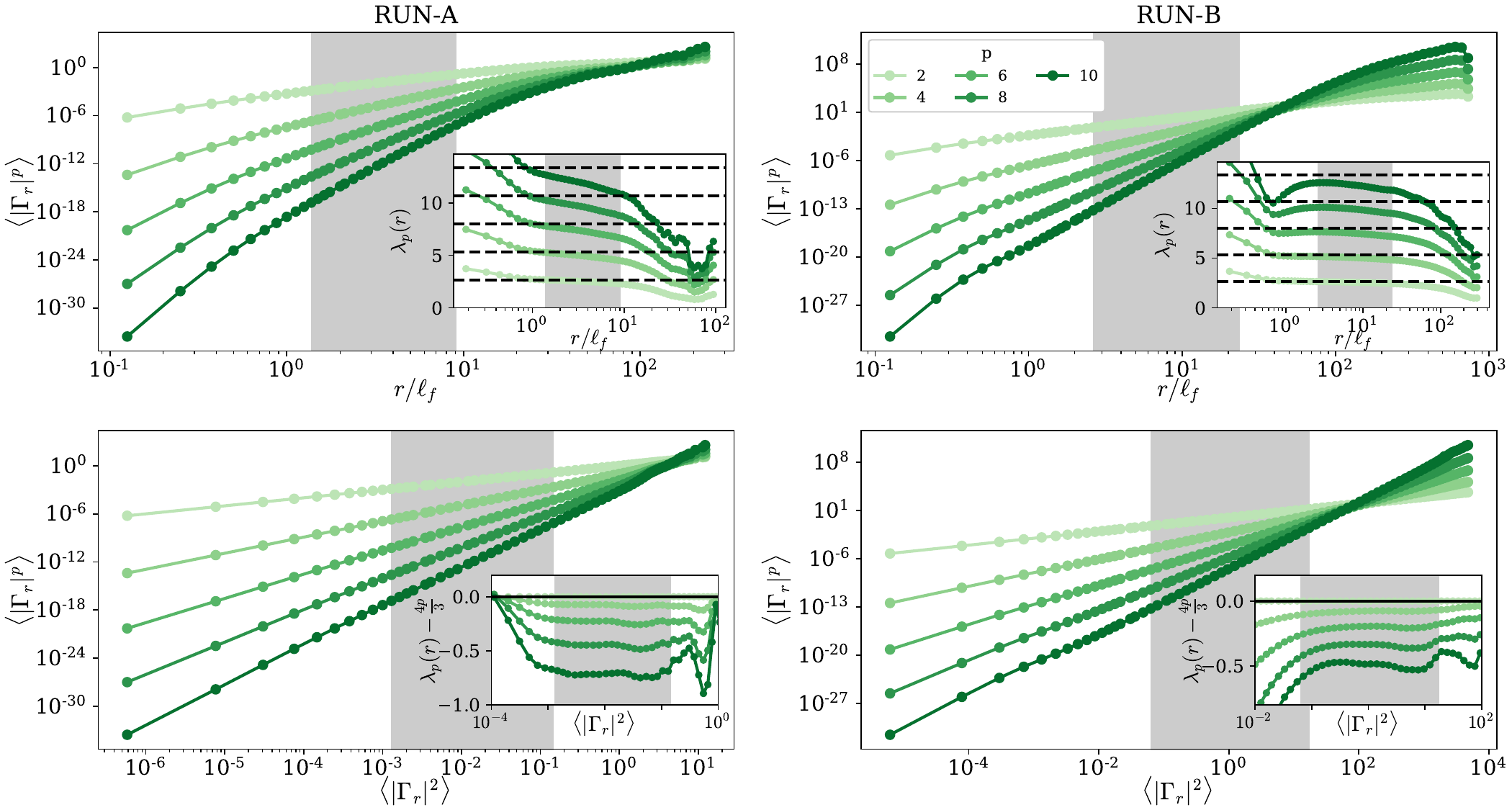}
  \caption[]{%
    Circulation moments up to $p=10$ in numerical simulations of 2D turbulence in the inverse energy cascade for (left column) RUN-A and (right column) RUN-B. Top panels show the circulation moments as a function of $r$, and the bottom panels show the ESS method. The shaded area indicates the inertial range. Dashed horizontal lines in the insets show the self-similar scaling $\lambda_p = 4p/3$. The top insets show the local slopes $\lambda_p(r) = \dd \log \langle \Gamma_r^p \rangle / \dd \log r$, and the bottom insets show the deviations from the self-similar scaling.   
    }
    \label{fig:circulation}
\end{figure*}

We now focus on the circulation statistics to see if there is a relation with the intermittency of TSFs.
We compute the velocity circulation around squared planar loops of size $r$ over the whole space. The calculations are performed in Fourier space to preserve the spectral precision from the simulations \cite{Muller2021}. 
The circulation moments $\langle |\Gamma_r|^p \rangle$ are shown in Fig.~\ref{fig:circulation}. 
The local slopes also exhibit poor scaling properties, which are improved by employing the ESS method with respect to the second-order moment $\langle |\Gamma_r|^2 \rangle$. The insets of the bottom panels in Fig.~\ref{fig:circulation} show the deviations of the local slopes with respect to the self-similar prediction $4p/3$. 

\begin{figure*}[h!]
  \centering
  \includegraphics[width=.99\textwidth]{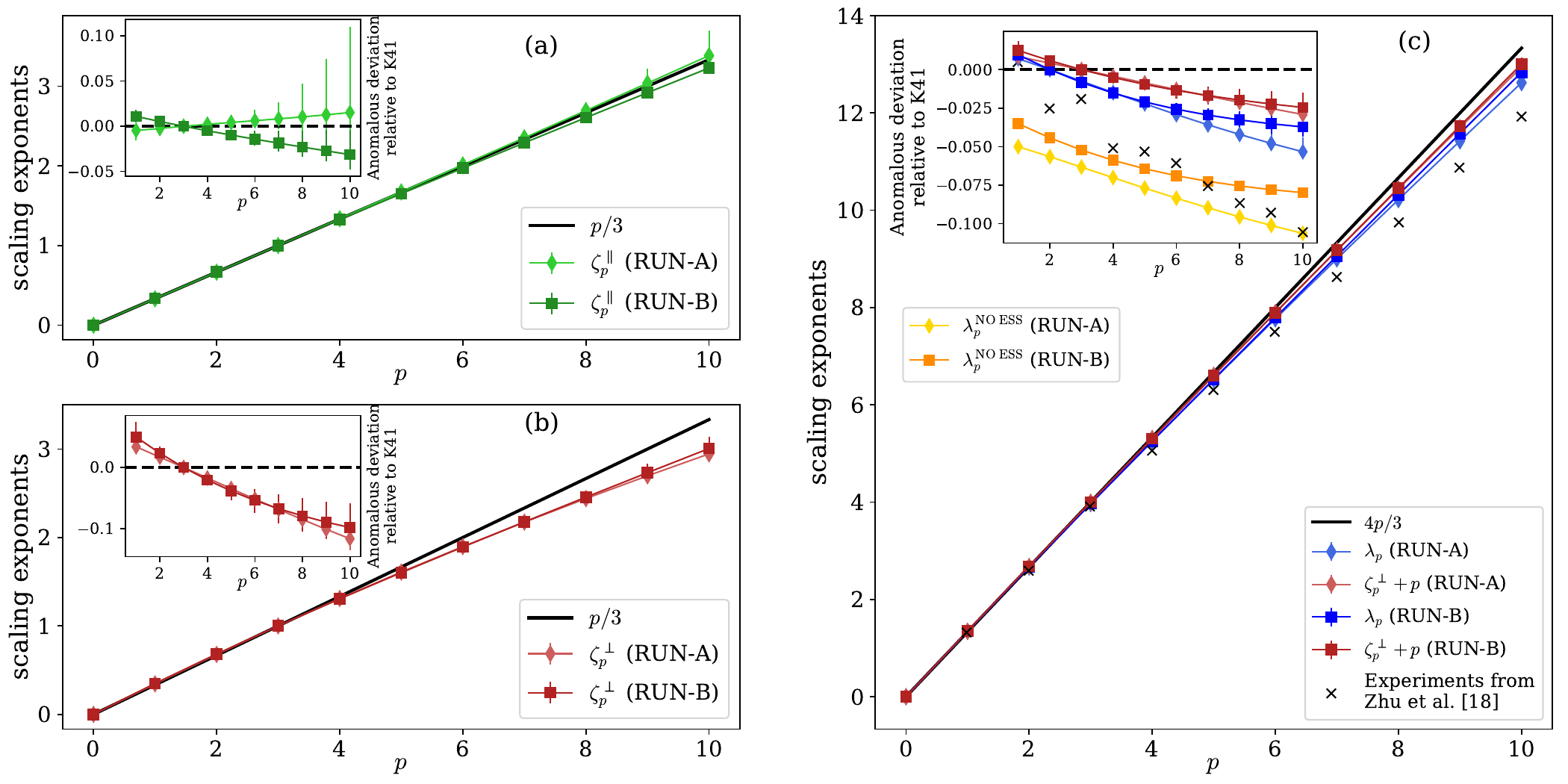}
  \caption[]{%
    Scaling exponents in the inverse energy cascade of 2D turbulence. Panels (a) and (b) show the longitudinal and transverse scaling exponents, respectively, for both numerical simulations up to $p=10$. Panel (c) shows the circulation exponents $\lambda_p$ compared to the transverse exponents $\zeta^\perp_p + p$. 
    The insets show the anomalous deviations with respect to the self-similar prediction $p/3$ for velocity increments and $4p/3$ for circulation defined in Eq.~\eqref{eq:relative_deviation}. Error bars are defined as the maximum and minimum values of the local slopes, which are not shown in the insets. Black crosses show experimental scaling exponents extracted from \cite{Zhu2023}. 
    }
    \label{fig:exponents}
\end{figure*}
Figure \ref{fig:exponents} summarizes the results of this work. 
The left panels show the longitudinal and transverse scaling exponents obtained from both databases using the ESS method, exhibiting a self-similar behavior for the former and an anomalous scaling for the latter. The error bars indicate the maximum and minimum values of the local slopes in the inertial range. 
The right panel shows the circulation scaling exponents $\lambda_p$ obtained from the ESS method, and we compare them with the transverse exponents via $\zeta_p^\perp + p$. We also include experimental data extracted from \citet{Zhu2023}, obtained without employing the ESS method. 
The insets show the anomalous deviations of the scaling exponents with respect to the self-similar predictions $\zeta_p^{\mathrm{ss}}=p/3$ and $\lambda_p^{\mathrm{ss}}=4p/3$, i.e. 
\begin{equation}
  \frac{\zeta_p^\parallel - \zeta_p^{\mathrm{ss}}}{\zeta_p^{\mathrm{ss}}},\quad\frac{\zeta_p^\perp - \zeta_p^{\mathrm{ss}}}{\zeta_p^{\mathrm{ss}}},\quad {\rm and}\,\,\frac{\lambda_p - \lambda_p^{\mathrm{ss}}}{\lambda_p^{\mathrm{ss}}}.
  \label{eq:relative_deviation}
\end{equation}
Despite the clear differences observed in the forcing scales between the two numerical simulations, the scaling exponents obtained from the ESS method in the inertial range are statistically equivalent. The different trend observed in the inset of Fig.\ref{fig:exponents}a might be due to statistical errors and it is less than $3\%$. Remarkably, both datasets lead to the same transverse anomalous scaling. Moreover, circulation scaling exponents measured in experiments coincide well with the ones measured in both simulations without using the ESS method.
These results show the robustness of the turbulent properties of the inverse cascade in 2D flows, at least when the ESS method is used. 

Starting from Eq.~\eqref{eq:circulation_increments} relating circulation and transverse velocity increments, it is possible to show using the H\"older inequality that (see Supplemental Material \cite{Supplemental})
\begin{equation}
  \lambda_p \leq \zeta_p^\perp + p.
  \label{eq:lambdap_zetap}
\end{equation}
This inequality implies that, if the TSFs are intermittent, i.e. $\zeta_p^\perp < p/3$ for $p>3$, then the circulation moments also are, as it follows that $\lambda_p < 4p/3$ for $p>3$. This relation is obeyed in our simulations, as shown in Fig.~\ref{fig:exponents}.c.
We remark that the derivation of Eq.~\eqref{eq:lambdap_zetap} is valid only in the limit $r/L_f \gg 1$ for the inverse energy cascade. In 3D turbulence, using data extracted from \citet{Iyer2019} and \citet{Iyer2020}, we observe that the inequality still holds, however, the demonstration of Eq.~\eqref{eq:lambdap_zetap} needs to be modified.

To conclude, we analyzed the statistical properties of two-dimensional turbulence from direct numerical simulations of the incompressible Navier-Stokes equations of two different databases. By forcing at small scales, a stationary inverse energy cascade is generated which exhibits well-known features, such as the $k^{-5/3}$ scaling law for the energy spectrum and a negative energy flux. The study of high-order moments structure functions confirmed the well-known self-similar behavior of longitudinal velocity increments but revealed anomalous deviations for the transverse ones. We argue that the intermittent behavior of this quantity explains the anomalous exponents of velocity circulation through the inequality \eqref{eq:lambdap_zetap}. 
For a future work we propose to investigate the inverse energy cascade in a dual cascade regime to study how non-local effects of the enstrophy cascade affect large-scale fluctuations. 
We also propose to study more realistic geophysical quasi-2D flows such as rotating or stratified turbulence. The characterization of structure functions and circulation moments in these systems could bring some insights into anisotropic effects and could be useful for atmospheric modeling. 

\section{Acknowledgments}
\begin{acknowledgments}
  We are grateful to Guido Boffetta for fruitful scientific discussions and for kindly sharing his 2D turbulence data. 
  This work was supported by the Agence Nationale de la Recherche through the project GIANTE ANR-18-CE30-0020-01. GK was also funded by the Simons Foundation project "Collaboration in Wave Turbulence" (award ID 651471)
  Computations were carried out at the Mésocentre SIGAMM hosted at the Observatoire de la Côte d'Azur.
\end{acknowledgments}

\appendix

\section{Appendix: Relation between circulation and transverse scaling exponents}

We consider the circulation $\Gamma_r(\svec_0)$ around a squared loop of size $r$, with one corner of the loop placed at $\svec_0 = (x_0, y_0)$. It follows that
\begin{equation}
  \langle|\Gamma_r|^p\rangle= \frac{1}{V} \int  |\Gamma_r(\svec_0)|^p\mathrm{d}\svec_0
   \leq \frac{1}{V} \int \left[\int_{x_0}^{x_0+r} | u_x(x, y_0+r) - u_x(x, y_0) | \dd x +  \int_{y_0}^{y_0+r} | u_y(x_0+r, y) - u_y(x_0,y) | \dd y\right]^p\mathrm{d}\svec_0,
  \label{eq:circulation_cauchy}
\end{equation}
where we applied the triangular inequality several times. Following a similar procedure to the one in \citet{Iyer2019}, we now apply the H\"older inequality for each of the two integrals in the right hand side. It leads to 
\begin{equation}
  \langle|\Gamma_r|^p\rangle\leq 
   \frac{1}{V} \int \left[ r^{1/q} \left(\int_{x_0}^{x_0+r} | u_x(x, y_0+r) - u_x(x, y_0) |^p \dd x \right)^{1/p} + r^{1/q} \left(\int_{y_0}^{y_0+r} | u_y(x_0+r, y) - u_y(x_0,y) |^p \dd y \right)^{1/p}  \right]^p\mathrm{d}\svec_0
  \label{eq:circulation_holder}
\end{equation}
with $p$ and $q$ satisfying $p^{-1} + q^{-1} = 1$ for $p,q > 1$. 

\begin{figure*}[h!]
  \centering
  \includegraphics[width=.5\textwidth]{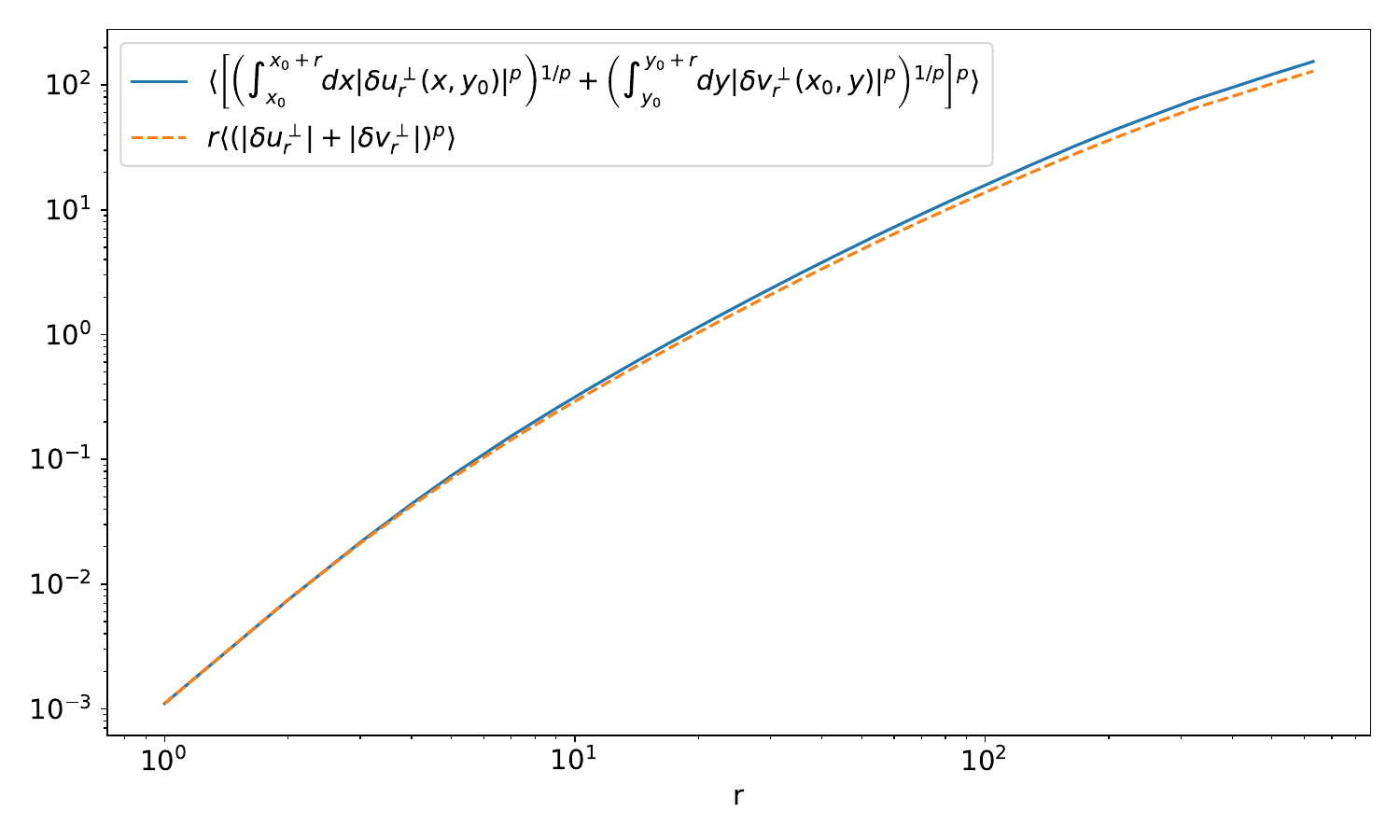}
  \caption[]{%
    Validation of the approximation performed between Eqs.\eqref{eq:circulation_holder} and \eqref{eq:circulation_approximation} for $p=2$, with $u = u_x$ and $v = u_y$. The angular brackets $\langle . \rangle$ indicate averaging in space. 
    }
    \label{fig:approximation}
\end{figure*}

For a sufficiently large Reynolds numbers, assuming homogeneity, isotropy, and at a fixed $r$ in the inertial range, we can approximate each inner integral by $r \langle |\delta u^\perp_r|^p \rangle = r S_p^\perp(r)$. Fig. \ref{fig:approximation} shows the validity of this approximation in the inertial range for RUN-A. 
The outer integral cancels out and we obtain
\begin{equation}
  \langle|\Gamma_r|^p\rangle \leq 2^p r^{p/q} r^{p/p} (S_p^{\perp})^{p/p}= 2^p r^p S_p^{\perp}(r).
  \label{eq:circulation_approximation}
\end{equation}

Finally, we use the fact that the circulation moments and TSFs follow the scaling properties $\langle |\Gamma_r|^p \rangle \sim (r/L_f)^{\lambda_p}$ and $S_p^{\perp} \sim (r/L_f)^{\zeta_p^\perp}$, with $L_f$ the forcing scale. 
For the inertial range of the inverse energy cascade in two-dimensional turbulence, we take the limit $r/L_f \gg 1$, so we obtain an inequality for the scaling exponents
\begin{equation}
  \lambda_p \leq \zeta_p^\perp + p.
  \label{eq:circulation_transverse_exponents}
\end{equation}

\bibliography{bibliography}

\begin{thebibliography}{25}%
\makeatletter
\providecommand \@ifxundefined [1]{%
 \@ifx{#1\undefined}
}%
\providecommand \@ifnum [1]{%
 \ifnum #1\expandafter \@firstoftwo
 \else \expandafter \@secondoftwo
 \fi
}%
\providecommand \@ifx [1]{%
 \ifx #1\expandafter \@firstoftwo
 \else \expandafter \@secondoftwo
 \fi
}%
\providecommand \natexlab [1]{#1}%
\providecommand \enquote  [1]{``#1''}%
\providecommand \bibnamefont  [1]{#1}%
\providecommand \bibfnamefont [1]{#1}%
\providecommand \citenamefont [1]{#1}%
\providecommand \href@noop [0]{\@secondoftwo}%
\providecommand \href [0]{\begingroup \@sanitize@url \@href}%
\providecommand \@href[1]{\@@startlink{#1}\@@href}%
\providecommand \@@href[1]{\endgroup#1\@@endlink}%
\providecommand \@sanitize@url [0]{\catcode `\\12\catcode `\$12\catcode `\&12\catcode `\#12\catcode `\^12\catcode `\_12\catcode `\%12\relax}%
\providecommand \@@startlink[1]{}%
\providecommand \@@endlink[0]{}%
\providecommand \url  [0]{\begingroup\@sanitize@url \@url }%
\providecommand \@url [1]{\endgroup\@href {#1}{\urlprefix }}%
\providecommand \urlprefix  [0]{URL }%
\providecommand \Eprint [0]{\href }%
\providecommand \doibase [0]{https://doi.org/}%
\providecommand \selectlanguage [0]{\@gobble}%
\providecommand \bibinfo  [0]{\@secondoftwo}%
\providecommand \bibfield  [0]{\@secondoftwo}%
\providecommand \translation [1]{[#1]}%
\providecommand \BibitemOpen [0]{}%
\providecommand \bibitemStop [0]{}%
\providecommand \bibitemNoStop [0]{.\EOS\space}%
\providecommand \EOS [0]{\spacefactor3000\relax}%
\providecommand \BibitemShut  [1]{\csname bibitem#1\endcsname}%
\let\auto@bib@innerbib\@empty
\bibitem [{\citenamefont {Kraichnan}(1967)}]{Kraichnan1967}%
  \BibitemOpen
  \bibfield  {author} {\bibinfo {author} {\bibfnamefont {R.~H.}\ \bibnamefont {Kraichnan}},\ }\bibfield  {title} {\bibinfo {title} {Inertial {{Ranges}} in {{Two-Dimensional Turbulence}}},\ }\href {https://doi.org/10.1063/1.1762301} {\bibfield  {journal} {\bibinfo  {journal} {Physics of Fluids}\ }\textbf {\bibinfo {volume} {10}},\ \bibinfo {pages} {1417} (\bibinfo {year} {1967})}\BibitemShut {NoStop}%
\bibitem [{\citenamefont {Leith}(1968)}]{Leith1968}%
  \BibitemOpen
  \bibfield  {author} {\bibinfo {author} {\bibfnamefont {C.~E.}\ \bibnamefont {Leith}},\ }\bibfield  {title} {\bibinfo {title} {Diffusion {{Approximation}} for {{Two-Dimensional Turbulence}}},\ }\href {https://doi.org/10.1063/1.1691968} {\bibfield  {journal} {\bibinfo  {journal} {Physics of Fluids}\ }\textbf {\bibinfo {volume} {11}},\ \bibinfo {pages} {671} (\bibinfo {year} {1968})}\BibitemShut {NoStop}%
\bibitem [{\citenamefont {Batchelor}(1969)}]{Batchelor1969}%
  \BibitemOpen
  \bibfield  {author} {\bibinfo {author} {\bibfnamefont {G.~K.}\ \bibnamefont {Batchelor}},\ }\bibfield  {title} {\bibinfo {title} {Computation of the {{Energy Spectrum}} in {{Homogeneous Two-Dimensional Turbulence}}},\ }\href {https://doi.org/10.1063/1.1692443} {\bibfield  {journal} {\bibinfo  {journal} {Physics of Fluids}\ }\textbf {\bibinfo {volume} {12}},\ \bibinfo {pages} {II} (\bibinfo {year} {1969})}\BibitemShut {NoStop}%
\bibitem [{\citenamefont {Davidson}(2013)}]{Davidson2013}%
  \BibitemOpen
  \bibfield  {author} {\bibinfo {author} {\bibfnamefont {P.~A.}\ \bibnamefont {Davidson}},\ }\href {https://doi.org/10.1017/CBO9781139208673} {\emph {\bibinfo {title} {Turbulence in {{Rotating}}, {{Stratified}} and {{Electrically Conducting Fluids}}}}}\ (\bibinfo  {publisher} {Cambridge University Press},\ \bibinfo {address} {Cambridge},\ \bibinfo {year} {2013})\BibitemShut {NoStop}%
\bibitem [{\citenamefont {Alexakis}\ and\ \citenamefont {Biferale}(2018)}]{Alexakis2018}%
  \BibitemOpen
  \bibfield  {author} {\bibinfo {author} {\bibfnamefont {A.}~\bibnamefont {Alexakis}}\ and\ \bibinfo {author} {\bibfnamefont {L.}~\bibnamefont {Biferale}},\ }\bibfield  {title} {\bibinfo {title} {Cascades and transitions in turbulent flows},\ }\href {https://doi.org/10.1016/j.physrep.2018.08.001} {\bibfield  {journal} {\bibinfo  {journal} {Physics Reports}\ }\textbf {\bibinfo {volume} {767--769}},\ \bibinfo {pages} {1} (\bibinfo {year} {2018})}\BibitemShut {NoStop}%
\bibitem [{\citenamefont {Marino}\ \emph {et~al.}(2013)\citenamefont {Marino}, \citenamefont {Mininni}, \citenamefont {Rosenberg},\ and\ \citenamefont {Pouquet}}]{Marino2013}%
  \BibitemOpen
  \bibfield  {author} {\bibinfo {author} {\bibfnamefont {R.}~\bibnamefont {Marino}}, \bibinfo {author} {\bibfnamefont {P.~D.}\ \bibnamefont {Mininni}}, \bibinfo {author} {\bibfnamefont {D.}~\bibnamefont {Rosenberg}},\ and\ \bibinfo {author} {\bibfnamefont {A.}~\bibnamefont {Pouquet}},\ }\bibfield  {title} {\bibinfo {title} {Inverse cascades in rotating stratified turbulence: {{Fast}} growth of large scales},\ }\href {https://doi.org/10.1209/0295-5075/102/44006} {\bibfield  {journal} {\bibinfo  {journal} {EPL (Europhysics Letters)}\ }\textbf {\bibinfo {volume} {102}},\ \bibinfo {pages} {44006} (\bibinfo {year} {2013})}\BibitemShut {NoStop}%
\bibitem [{\citenamefont {Young}\ and\ \citenamefont {Read}(2017)}]{Young2017}%
  \BibitemOpen
  \bibfield  {author} {\bibinfo {author} {\bibfnamefont {R.~M.~B.}\ \bibnamefont {Young}}\ and\ \bibinfo {author} {\bibfnamefont {P.~L.}\ \bibnamefont {Read}},\ }\bibfield  {title} {\bibinfo {title} {Forward and inverse kinetic energy cascades in {{Jupiter}}'s turbulent weather layer},\ }\href {https://doi.org/10.1038/nphys4227} {\bibfield  {journal} {\bibinfo  {journal} {Nature Physics}\ }\textbf {\bibinfo {volume} {13}},\ \bibinfo {pages} {1135} (\bibinfo {year} {2017})}\BibitemShut {NoStop}%
\bibitem [{\citenamefont {Frisch}(1995)}]{Frisch1995}%
  \BibitemOpen
  \bibfield  {author} {\bibinfo {author} {\bibfnamefont {U.}~\bibnamefont {Frisch}},\ }\href {https://doi.org/10.1017/CBO9781139170666} {\emph {\bibinfo {title} {Turbulence: {{The Legacy}} of {{A}}.{{N}}. {{Kolmogorov}}}}},\ \bibinfo {edition} {1st}\ ed.\ (\bibinfo  {publisher} {Cambridge University Press},\ \bibinfo {address} {Cambridge},\ \bibinfo {year} {1995})\BibitemShut {NoStop}%
\bibitem [{\citenamefont {Paret}\ and\ \citenamefont {Tabeling}(1998)}]{Paret1998}%
  \BibitemOpen
  \bibfield  {author} {\bibinfo {author} {\bibfnamefont {J.}~\bibnamefont {Paret}}\ and\ \bibinfo {author} {\bibfnamefont {P.}~\bibnamefont {Tabeling}},\ }\bibfield  {title} {\bibinfo {title} {Intermittency in the two-dimensional inverse cascade of energy: {{Experimental}} observations},\ }\href {https://doi.org/10.1063/1.869840} {\bibfield  {journal} {\bibinfo  {journal} {Physics of Fluids}\ }\textbf {\bibinfo {volume} {10}},\ \bibinfo {pages} {3126} (\bibinfo {year} {1998})}\BibitemShut {NoStop}%
\bibitem [{\citenamefont {Boffetta}\ \emph {et~al.}(2000)\citenamefont {Boffetta}, \citenamefont {Celani},\ and\ \citenamefont {Vergassola}}]{Boffetta2000}%
  \BibitemOpen
  \bibfield  {author} {\bibinfo {author} {\bibfnamefont {G.}~\bibnamefont {Boffetta}}, \bibinfo {author} {\bibfnamefont {A.}~\bibnamefont {Celani}},\ and\ \bibinfo {author} {\bibfnamefont {M.}~\bibnamefont {Vergassola}},\ }\bibfield  {title} {\bibinfo {title} {Inverse energy cascade in two-dimensional turbulence: {{Deviations}} from {{Gaussian}} behavior},\ }\href {https://doi.org/10.1103/PhysRevE.61.R29} {\bibfield  {journal} {\bibinfo  {journal} {Physical Review E}\ }\textbf {\bibinfo {volume} {61}},\ \bibinfo {pages} {R29} (\bibinfo {year} {2000})}\BibitemShut {NoStop}%
\bibitem [{\citenamefont {Iyer}\ \emph {et~al.}(2019)\citenamefont {Iyer}, \citenamefont {Sreenivasan},\ and\ \citenamefont {Yeung}}]{Iyer2019}%
  \BibitemOpen
  \bibfield  {author} {\bibinfo {author} {\bibfnamefont {K.~P.}\ \bibnamefont {Iyer}}, \bibinfo {author} {\bibfnamefont {K.~R.}\ \bibnamefont {Sreenivasan}},\ and\ \bibinfo {author} {\bibfnamefont {P.~K.}\ \bibnamefont {Yeung}},\ }\bibfield  {title} {\bibinfo {title} {Circulation in {{High Reynolds Number Isotropic Turbulence}} is a {{Bifractal}}},\ }\href {https://doi.org/10.1103/PhysRevX.9.041006} {\bibfield  {journal} {\bibinfo  {journal} {Physical Review X}\ }\textbf {\bibinfo {volume} {9}},\ \bibinfo {pages} {041006} (\bibinfo {year} {2019})}\BibitemShut {NoStop}%
\bibitem [{\citenamefont {M{\"u}ller}\ \emph {et~al.}(2021)\citenamefont {M{\"u}ller}, \citenamefont {Polanco},\ and\ \citenamefont {Krstulovic}}]{Muller2021}%
  \BibitemOpen
  \bibfield  {author} {\bibinfo {author} {\bibfnamefont {N.~P.}\ \bibnamefont {M{\"u}ller}}, \bibinfo {author} {\bibfnamefont {J.~I.}\ \bibnamefont {Polanco}},\ and\ \bibinfo {author} {\bibfnamefont {G.}~\bibnamefont {Krstulovic}},\ }\bibfield  {title} {\bibinfo {title} {Intermittency of {{Velocity Circulation}} in {{Quantum Turbulence}}},\ }\href {https://doi.org/10.1103/PhysRevX.11.011053} {\bibfield  {journal} {\bibinfo  {journal} {Physical Review X}\ }\textbf {\bibinfo {volume} {11}},\ \bibinfo {pages} {011053} (\bibinfo {year} {2021})}\BibitemShut {NoStop}%
\bibitem [{\citenamefont {Polanco}\ \emph {et~al.}(2021)\citenamefont {Polanco}, \citenamefont {M{\"u}ller},\ and\ \citenamefont {Krstulovic}}]{Polanco2021}%
  \BibitemOpen
  \bibfield  {author} {\bibinfo {author} {\bibfnamefont {J.~I.}\ \bibnamefont {Polanco}}, \bibinfo {author} {\bibfnamefont {N.~P.}\ \bibnamefont {M{\"u}ller}},\ and\ \bibinfo {author} {\bibfnamefont {G.}~\bibnamefont {Krstulovic}},\ }\bibfield  {title} {\bibinfo {title} {Vortex clustering, polarisation and circulation intermittency in classical and quantum turbulence},\ }\href {https://doi.org/10.1038/s41467-021-27382-6} {\bibfield  {journal} {\bibinfo  {journal} {Nature Communications}\ }\textbf {\bibinfo {volume} {12}},\ \bibinfo {pages} {7090} (\bibinfo {year} {2021})}\BibitemShut {NoStop}%
\bibitem [{\citenamefont {M{\"u}ller}\ \emph {et~al.}(2022)\citenamefont {M{\"u}ller}, \citenamefont {Tang}, \citenamefont {Guo},\ and\ \citenamefont {Krstulovic}}]{Muller2022a}%
  \BibitemOpen
  \bibfield  {author} {\bibinfo {author} {\bibfnamefont {N.~P.}\ \bibnamefont {M{\"u}ller}}, \bibinfo {author} {\bibfnamefont {Y.}~\bibnamefont {Tang}}, \bibinfo {author} {\bibfnamefont {W.}~\bibnamefont {Guo}},\ and\ \bibinfo {author} {\bibfnamefont {G.}~\bibnamefont {Krstulovic}},\ }\bibfield  {title} {\bibinfo {title} {Velocity circulation intermittency in finite-temperature turbulent superfluid helium},\ }\href {https://doi.org/10.1103/PhysRevFluids.7.104604} {\bibfield  {journal} {\bibinfo  {journal} {Physical Review Fluids}\ }\textbf {\bibinfo {volume} {7}},\ \bibinfo {pages} {104604} (\bibinfo {year} {2022})}\BibitemShut {NoStop}%
\bibitem [{\citenamefont {Migdal}(2020)}]{Migdal2020}%
  \BibitemOpen
  \bibfield  {author} {\bibinfo {author} {\bibfnamefont {A.}~\bibnamefont {Migdal}},\ }\bibfield  {title} {\bibinfo {title} {Clebsch confinement and instantons in turbulence},\ }\href {https://doi.org/10.1142/S0217751X20300185} {\bibfield  {journal} {\bibinfo  {journal} {International Journal of Modern Physics A}\ }\textbf {\bibinfo {volume} {35}},\ \bibinfo {pages} {2030018} (\bibinfo {year} {2020})}\BibitemShut {NoStop}%
\bibitem [{\citenamefont {Apolin{\'a}rio}\ \emph {et~al.}(2020)\citenamefont {Apolin{\'a}rio}, \citenamefont {Moriconi}, \citenamefont {Pereira},\ and\ \citenamefont {Valad{\~a}o}}]{Apolinario2020}%
  \BibitemOpen
  \bibfield  {author} {\bibinfo {author} {\bibfnamefont {G.~B.}\ \bibnamefont {Apolin{\'a}rio}}, \bibinfo {author} {\bibfnamefont {L.}~\bibnamefont {Moriconi}}, \bibinfo {author} {\bibfnamefont {R.~M.}\ \bibnamefont {Pereira}},\ and\ \bibinfo {author} {\bibfnamefont {V.~J.}\ \bibnamefont {Valad{\~a}o}},\ }\bibfield  {title} {\bibinfo {title} {Vortex gas modeling of turbulent circulation statistics},\ }\href {https://doi.org/10.1103/PhysRevE.102.041102} {\bibfield  {journal} {\bibinfo  {journal} {Physical Review E}\ }\textbf {\bibinfo {volume} {102}},\ \bibinfo {pages} {041102(R)} (\bibinfo {year} {2020})}\BibitemShut {NoStop}%
\bibitem [{\citenamefont {Moriconi}\ \emph {et~al.}(2024)\citenamefont {Moriconi}, \citenamefont {Pereira},\ and\ \citenamefont {Valad{\~a}o}}]{Moriconi2024}%
  \BibitemOpen
  \bibfield  {author} {\bibinfo {author} {\bibfnamefont {L.}~\bibnamefont {Moriconi}}, \bibinfo {author} {\bibfnamefont {R.~M.}\ \bibnamefont {Pereira}},\ and\ \bibinfo {author} {\bibfnamefont {V.~J.}\ \bibnamefont {Valad{\~a}o}},\ }\bibfield  {title} {\bibinfo {title} {Vortex polarization and circulation statistics in isotropic turbulence},\ }\href {https://doi.org/10.1103/PhysRevE.109.045106} {\bibfield  {journal} {\bibinfo  {journal} {Physical Review E}\ }\textbf {\bibinfo {volume} {109}},\ \bibinfo {pages} {045106} (\bibinfo {year} {2024})}\BibitemShut {NoStop}%
\bibitem [{\citenamefont {Zhu}\ \emph {et~al.}(2023)\citenamefont {Zhu}, \citenamefont {Xie},\ and\ \citenamefont {Xia}}]{Zhu2023}%
  \BibitemOpen
  \bibfield  {author} {\bibinfo {author} {\bibfnamefont {H.-Y.}\ \bibnamefont {Zhu}}, \bibinfo {author} {\bibfnamefont {J.-H.}\ \bibnamefont {Xie}},\ and\ \bibinfo {author} {\bibfnamefont {K.-Q.}\ \bibnamefont {Xia}},\ }\bibfield  {title} {\bibinfo {title} {Circulation in {{Quasi-2D Turbulence}}: {{Experimental Observation}} of the {{Area Rule}} and {{Bifractality}}},\ }\href {https://doi.org/10.1103/PhysRevLett.130.214001} {\bibfield  {journal} {\bibinfo  {journal} {Physical Review Letters}\ }\textbf {\bibinfo {volume} {130}},\ \bibinfo {pages} {214001} (\bibinfo {year} {2023})}\BibitemShut {NoStop}%
\bibitem [{\citenamefont {M{\"u}ller}\ and\ \citenamefont {Krstulovic}(2024)}]{Muller2024}%
  \BibitemOpen
  \bibfield  {author} {\bibinfo {author} {\bibfnamefont {N.~P.}\ \bibnamefont {M{\"u}ller}}\ and\ \bibinfo {author} {\bibfnamefont {G.}~\bibnamefont {Krstulovic}},\ }\bibfield  {title} {\bibinfo {title} {Exploring the {{Equivalence}} between {{Two-Dimensional Classical}} and {{Quantum Turbulence}} through {{Velocity Circulation Statistics}}},\ }\href {https://doi.org/10.1103/PhysRevLett.132.094002} {\bibfield  {journal} {\bibinfo  {journal} {Physical Review Letters}\ }\textbf {\bibinfo {volume} {132}},\ \bibinfo {pages} {094002} (\bibinfo {year} {2024})}\BibitemShut {NoStop}%
\bibitem [{Sup()}]{Supplemental}%
  \BibitemOpen
  \href@noop {} {}\bibinfo {note} {See Supplemental Material at URL-will-be-inserted-by-publisher for details on the analytical derivation of Eq.~\eqref{eq:lambdap_zetap}.}\BibitemShut {Stop}%
\bibitem [{\citenamefont {Chen}\ \emph {et~al.}(1997)\citenamefont {Chen}, \citenamefont {Sreenivasan}, \citenamefont {Nelkin},\ and\ \citenamefont {Cao}}]{Chen1997}%
  \BibitemOpen
  \bibfield  {author} {\bibinfo {author} {\bibfnamefont {S.}~\bibnamefont {Chen}}, \bibinfo {author} {\bibfnamefont {K.~R.}\ \bibnamefont {Sreenivasan}}, \bibinfo {author} {\bibfnamefont {M.}~\bibnamefont {Nelkin}},\ and\ \bibinfo {author} {\bibfnamefont {N.}~\bibnamefont {Cao}},\ }\bibfield  {title} {\bibinfo {title} {Refined {{Similarity Hypothesis}} for {{Transverse Structure Functions}} in {{Fluid Turbulence}}},\ }\href {https://doi.org/10.1103/PhysRevLett.79.2253} {\bibfield  {journal} {\bibinfo  {journal} {Physical Review Letters}\ }\textbf {\bibinfo {volume} {79}},\ \bibinfo {pages} {2253} (\bibinfo {year} {1997})}\BibitemShut {NoStop}%
\bibitem [{\citenamefont {Iyer}\ \emph {et~al.}(2020)\citenamefont {Iyer}, \citenamefont {Sreenivasan},\ and\ \citenamefont {Yeung}}]{Iyer2020}%
  \BibitemOpen
  \bibfield  {author} {\bibinfo {author} {\bibfnamefont {K.~P.}\ \bibnamefont {Iyer}}, \bibinfo {author} {\bibfnamefont {K.~R.}\ \bibnamefont {Sreenivasan}},\ and\ \bibinfo {author} {\bibfnamefont {P.~K.}\ \bibnamefont {Yeung}},\ }\bibfield  {title} {\bibinfo {title} {Scaling exponents saturate in three-dimensional isotropic turbulence},\ }\href {https://doi.org/10.1103/PhysRevFluids.5.054605} {\bibfield  {journal} {\bibinfo  {journal} {Physical Review Fluids}\ }\textbf {\bibinfo {volume} {5}},\ \bibinfo {pages} {054605} (\bibinfo {year} {2020})}\BibitemShut {NoStop}%
\bibitem [{\citenamefont {Boffetta}\ and\ \citenamefont {Ecke}(2012)}]{Boffetta2012}%
  \BibitemOpen
  \bibfield  {author} {\bibinfo {author} {\bibfnamefont {G.}~\bibnamefont {Boffetta}}\ and\ \bibinfo {author} {\bibfnamefont {R.~E.}\ \bibnamefont {Ecke}},\ }\bibfield  {title} {\bibinfo {title} {Two-{{Dimensional Turbulence}}},\ }\href {https://doi.org/10.1146/annurev-fluid-120710-101240} {\bibfield  {journal} {\bibinfo  {journal} {Annual Review of Fluid Mechanics}\ }\textbf {\bibinfo {volume} {44}},\ \bibinfo {pages} {427} (\bibinfo {year} {2012})}\BibitemShut {NoStop}%
\bibitem [{\citenamefont {Boffetta}(2007)}]{Boffetta2007}%
  \BibitemOpen
  \bibfield  {author} {\bibinfo {author} {\bibfnamefont {G.}~\bibnamefont {Boffetta}},\ }\bibfield  {title} {\bibinfo {title} {Energy and enstrophy fluxes in the double cascade of two-dimensional turbulence},\ }\href {https://doi.org/10.1017/S0022112007008014} {\bibfield  {journal} {\bibinfo  {journal} {Journal of Fluid Mechanics}\ }\textbf {\bibinfo {volume} {589}},\ \bibinfo {pages} {253} (\bibinfo {year} {2007})}\BibitemShut {NoStop}%
\bibitem [{\citenamefont {Benzi}\ \emph {et~al.}(1993)\citenamefont {Benzi}, \citenamefont {Ciliberto}, \citenamefont {Tripiccione}, \citenamefont {Baudet}, \citenamefont {Massaioli},\ and\ \citenamefont {Succi}}]{Benzi1993}%
  \BibitemOpen
  \bibfield  {author} {\bibinfo {author} {\bibfnamefont {R.}~\bibnamefont {Benzi}}, \bibinfo {author} {\bibfnamefont {S.}~\bibnamefont {Ciliberto}}, \bibinfo {author} {\bibfnamefont {R.}~\bibnamefont {Tripiccione}}, \bibinfo {author} {\bibfnamefont {C.}~\bibnamefont {Baudet}}, \bibinfo {author} {\bibfnamefont {F.}~\bibnamefont {Massaioli}},\ and\ \bibinfo {author} {\bibfnamefont {S.}~\bibnamefont {Succi}},\ }\bibfield  {title} {\bibinfo {title} {Extended self-similarity in turbulent flows},\ }\href@noop {} {\bibfield  {journal} {\bibinfo  {journal} {Physical Review E}\ }\textbf {\bibinfo {volume} {48}},\ \bibinfo {pages} {R29} (\bibinfo {year} {1993})}\BibitemShut {NoStop}%
\end{thebibliography}%

\end{document}